\documentclass[11pt,twoside]{article}

\usepackage{asp2014}
\resetcounters

\begin{document}

\title{Tidal Disruption Events}

\author{Sjoert~van~Velzen,$^{1}$ Geoffrey C. Bower,$^{2}$ and Brian D. Metzger$^{3}$}
\affil{$^1$New York University, New York, NY 10003, USA; \email{sjoert@nyu.edu}}
\affil{$^2$Academia Sinica Institute of Astronomy and Astrophysics, Hilo, HI 96720, USA}
\affil{$^3$Columbia Astrophysics Laboratory, New York, NY 10027, USA} 

\paperauthor{Sjoert~van~Velzen}{sjoert@nyu.edu}{0000-0002-3859-8074}{New York University}{CCPP}{NY}{NY}{10003}{USA}
\paperauthor{Geoffrey~C.~Bower}{gbower@asiaa.sinica.edu.tw}{}{Academia Sinica}{Institute of Astronomy and Astrophysics}{Hilo}{HI}{96720}{USA}
\paperauthor{Brian~D.~Metzger}{bmetzger@phys.columbia.edu}{}{Columbia University}{Columbia Astrophysics Laboratory}{NY}{NY}{10027}{USA}

\begin{abstract}
The tidal disruption and subsequent accretion of a star by a supermassive black hole can be used as a laboratory to study the physics of relativistic jets. The ngVLA is the only planned instrument that can both discover and characterize a large number of these short-lived radio sources. In particular the high-frequency capabilities of the ngVLA enable this important leap forward. Multi-frequency radio follow-up observations ($3-100$\,GHz) of tidal disruption events found in optical or X-ray surveys will provide a measurement of the jet efficiency as a function of black hole spin, thus enabling a direct test of the prediction that relativistic jets require high spin. Hundreds of tidal disruption jets will be discovered in a blind ngVLA survey for radio transients. By including VLBI observations with the ngVLA Long Baseline Array, we can resolve some of these sources, obtaining a robust measurement of the jet launch date and the magnetic field strength. From the thermal emission of the tidal disruption flare we can measure the accretion rate at this launch date, thus providing another unique opportunity to identify the conditions that lead to jet production.
\end{abstract}

\section{Introduction: Thermal and Non-Thermal Emission}
A rare glimpse into the properties of normally quiescent massive black holes is afforded when a star passes sufficiently close that it is torn apart by the black hole's tidal gravitational field \citep{Hills75}.  The process of disruption leaves a significant fraction of the shredded star gravitationally bound to the black hole \citep{Rees88,EvansKochanek89} and this stellar debris has been predicted to power a thermal flare at optical, UV, and X-ray wavelengths. Over a dozen such thermal tidal disruption flare (TDF) candidates have now been identified (\citealt{Komossa15,vanVelzen18} and references therein). Based on the current detection rate in optical surveys, we expect that the Large Synoptic Survey Telescope (LSST) will yield thousands of thermal TDFs per year \citep{Gezari09,vanVelzen10}.  

Radio follow-up observations of thermal TDFs have mostly yielded non-detections \citep{Bower13,vanVelzen12b,Arcavi14}. Only two low-redshift thermal TDFs ($z<0.02$) have been detected in radio follow-up observations  \citep{vanVelzen16,Alexander16,Alexander17}. The origin of this radio emission is long debated, both a jet \citep{vanVelzen11}, a disk-wind \citep{Alexander16}, or unbound stellar debris \citep{Krolik16} have been proposed. For one event (ASASSN-14li; \citealt{Holoien16}), the detection of a cross-correlation between the X-ray and radio light curves provides evidence that the radio emission is produced internal to a moderately relativistic jet \citep{PashamvanVelzen17}.

While most known TDFs are dominated by thermal emission, a second class of events is discovered via non-thermal $\gamma$-ray emission. The most famous example in this class is Swift~J1644+57 \citep{Bloom11,Levan11,Burrows11,Zauderer11}. This source was characterized by powerful ($\nu L_{\nu}\sim 10^{47}$~erg\,s\,$^{-1}$) hard X-ray emission.  The long duration of Swift J1644+57 and a position coincident with the nucleus of a previously quiescent galaxy led to the conclusion that it was powered by rapid accretion onto the central black hole following a stellar disruption.  The rapid X-ray variability suggested an origin internal to a jet that is relativistically beaming its radiation along our line of sight, similar to the blazar geometry of normal active galactic nuclei \citep{Bloom11}.  

\begin{figure}[t!]
\begin{center}
\includegraphics[width=0.8\textwidth]{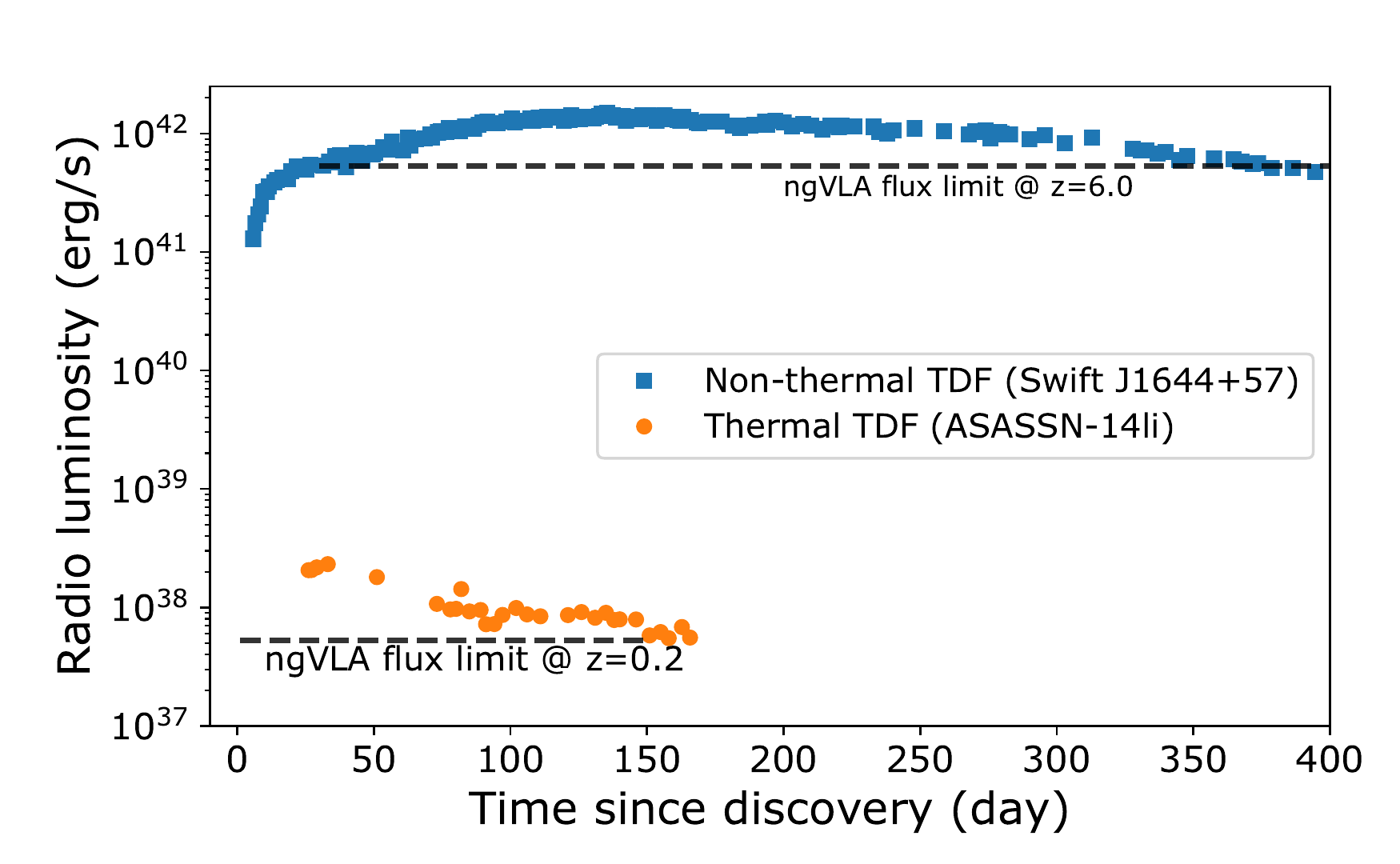} 
\end{center}
\caption{Radio light curve (15~GHz) of two archetype radio-emitting TDFs. The dashed line indicates a 5$\sigma$ detection threshold of a potential ngVLA transient survey using 10~minute snapshots at 8~GHz (Band~2) to a reach an image rms of $1~\mu$Jy/beam.  If most thermal TDFs are accompanied by radio emission similar to ASASSN-14li, a blind ngVLA survey covering 300~deg$^2$ should detect $\sim 10$ of these events per year, the expected detection rate of events similar to Swift~J1644+57 is an order of magnitude higher.}\label{fig:radiolc}
\end{figure}

Swift~J1644+57 was also characterized by luminous synchrotron radio emission, that brightened gradually over the course of several months \citep{Berger12,Zauderer13}, as shown in Figure \ref{fig:radiolc}. This radio emission resulted from the shock interaction between the stellar tidal disruption jet and the external gas surrounding the black hole \citep{Giannios11}, similar to a gamma-ray burst afterglow.  Two more jetted tidal disruption events with similar X-ray and radio properties to Swift J1644+57 have been found \citep{Cenko12,Pasham15,Brown15}.

\section{Using Tidal Disruption Events to Study the Physics of Jet Launching}
When a central object gathers mass through an accretion disk, jets are ubiquitous. Jets transport energy away from the disk, thus greatly enhancing the scale at which the accreting object (neutron star, black hole, or protostar) can influence its environment. As discussed in detail in earlier chapters of this book (e.g., Nyland et al. 2018, this volume), jets from supermassive black holes can provide strong negative feedback onto their host galaxy. However, adding jet mode feedback to cosmological galaxy simulations \citep[e.g.,][]{Vogelsberger14} requires a key ingredient that is currently missing: the jet duty cycle and lifetime. 
If powerful, long-lived jets can only be produced for a small fraction of active galactic nuclei (AGN), jet feedback will be limited to the subclass of galaxies that host these black holes. On the other hand, AGN may experience multiple cycles of jet production, implying a higher efficiency of inter-galaxy jet feedback. These two scenarios, long-lived versus intermittent jets, broadly correspond to two opposing theories of jet production: the spin paradigm and the accretion paradigm.

In the spin paradigm, jets are powered by the rotational energy of the black hole, which can be tapped using the magnetic field lines that thread the black hole horizon \citep{BlandfordZnajek77}. 
Only rapidly spinning black holes are observed to be radio loud. Alternatively, the low number of radio-loud quasars \citep[e.g.,][]{Kellermann89} can be explained by intermittent jet production due to a state change of the accretion disk \citep[][]{Falcke04,Koerding06}. This accretion-paradigm is motivated by observations of stellar mass black holes in X-ray binaries, where jet production triggered by a state change of the accretion disk can be observed directly \citep*{Fender04}. 

The current body of observations of AGN is not sufficient to solve the problem of jet launching because state changes of these systems are nearly\footnote{The so-called changing look AGN \citep[e.g.,][]{LaMassa15} could present an exception to normal AGN variability \citep{Graham17} and can be used to study the jet-disk connection over $\sim 10$~yr timescales \citep[e.g.,][]{Koay16}.} impossible to observe directly due to the long timescales associated with the feeding of AGN.  A tidal disruption event provides an opportunity to observe jets from supermassive black holes in real time. Since the fallback rate onto the black hole changes from super-Eddington to sub-Eddington in a few months to years, we may expect jet production due to state changes of the accretion disk \citep{Giannios11,vanVelzen11}. However our current radio observations of tidal disruption are not yet sufficient to make unambiguous inference on jet production. First of all, the number of radio-detected sources is very small. Second, the nature of radio emission from thermal TDFs is unclear. And third, the event rate of the non-thermal TDFs with powerful jets is unknown. 

The ngVLA will overcome all these limitations at once. A key element that makes this possible is the increased sensitivity at high-frequency ($\nu > 3$~GHz). Due to synchrotron self-absorption, TDF jets -- like most synchrotron-powered transients -- are very faint at low radio frequencies. Only when the source grows large enough, it becomes bright at $\nu \sim 1$~GHz. But at this large size, the typical timescale of the radio emission is long ($\sim 1$~yr); since this timescale is similar to the cadence of most surveys, the amplitude of most synchrotron transients observed at low frequencies is small. Finally, with the high angular resolution of the ngVLA Long Baseline Array we can resolve radio emission from TDFs and thus watch these newborn jets grow with time. 

In the next two subsections we discuss how the ngVLA can revolutionize TDF research using two different modes of observation: follow-up and discovery. 

\subsection{Follow-up of TDFs with the ngVLA}
Radio follow-up observations with current facilities have only yielded detections for low-redshift TDFs ($z<0.05$). Current optical surveys for thermal TDFs are nearly complete for these nearby events, finding about one per year. While the LSST and other optical or X-ray surveys will increase the detection rate of thermal TDFs to thousands per year, the  majority of these will be far too distant to be detectable with Jansky~VLA follow-up observations. This implies the detection rate of radio-emitting thermal TDFs will fall behind the overall detection rate of TDFs.

A tenfold increase to the sensitivity of the Jansky~VLA will push the maximum redshift for detection (within 10~min of integration time per source) of radio emission from thermal TDFs to $z=0.2$. The order of magnitude increase to the detection rate will enable measurements of the incidence of radio emission as a function of host galaxy properties. Most importantly, we can measure the jet production efficiency as a function of black hole spin. Disruptions by black holes more massive than $10^{8}\, M_{\odot}$ are visible only when the black hole is spinning \citep{Stone18}. Such events are rare, but the first TDF candidate from a black hole above this critical mass was recently detected \citep{Leloudas16,Margutti17,Kruhler18}---however see \citet{Dong16} for different interpretation. To conclude, ngVLA follow-up observations of thermal TDFs will yield a measurement of the radio luminosity as a function of black hole spin, thus providing a crucial test of the spin paradigm of jet production. 

\subsection{Discovery of TDFs in Blind ngVLA surveys}
Thanks to their high luminosity, radio emission from non-thermal TDF jets (such as Swift~J1644+57) can be detected by current radio surveys well into the dawn of galaxy formation \citep*{Metzger15}. As demonstrated in \citet{vanVelzen12b}, the detection rate of these events in surveys for radio transients can be estimated using the observed light curve of Swift~J1644+57. This estimate requires the volumetric event rate, which can in turn be estimated from the observed rate of on-axis tidal disruption jets ($\sim 10^{-2}\,{\rm Gpc}^{-3}{\rm yr}^{-1}$, based on the {\it Swift} data; \citealt{Brown15,Levan16}) and applying a beaming correction ($\delta$) to obtain the isotropic rate. The beaming correction can both be estimated from the requirement that the intrinsic X-ray luminosity should not exceed the Eddington limit, or by modeling the radio light curves. Both of these approaches yield $\delta\sim 10^{2}$, hence the volumetric rate of events similar to Swift~J1644+57 can be estimated to be $\sim 1\,{\rm Gpc}^{-3}{\rm yr}^{-1}$. 

Using a model to predict the off-axis light curve of events similar to Swift~J1644+57, the resulting detection rate of off-axis TDF jets in VLASS is about 10 per year. However the cadence of the VLASS is low ($\approx 16$~months), hence most sources will be detected after maximum light. The cadence of surveys at $\approx 1$ GHz by SKA precursors, such as VAST \citep{Murphy13} and ThunderKAT \citep{Fender17}, is higher. The detection rate for off-axis TDF jets in these surveys is also about $10$ per year \citep{Metzger15}. However, due to the very shallow amplitude of the 1~GHz light curves, the identification of these events using only SKA data will be difficult; coordinated  observations at higher frequencies (X-ray or optical) will be required \citep{Donnarumma15,Donnarumma15a}. Follow-up observations of SKA-discovered TDFs with the VLA or ALMA are difficult because the emission at these higher radio frequencies fades within months to weeks, while the emission at 1~GHz is delayed by about a year. Therefore most TDFs candidates found in SKA surveys will have monochromatic radio light curves.

The high frequency capabilities for the ngVLA will enable both the discovery {\it and} characterization of TDFs in blind radio surveys. Using observations from 3 to 100 GHz, we can measure the total energy of the jet and estimate the expansion velocity. By comparing the event rate from this blind radio survey to the observed rate of TDFs found by {\it Swift} we obtaind the beaming factor ($\delta$). The detection of a low beaming factor will imply a super-Eddington luminosity for the hard X-ray, as suggested in some models of Swift~J1644+57 \citep{Kara16}. Scaling from estimates by \citet{Metzger15} for the number of TDFs in VLASS (using $10^{4}\,{\rm deg}^{2}$ at an rms of $0.1$~mJy/beam), we can expect $\sim 10^{2}$ events per year for an ngVLA transient survey in Band 1 ($3$~GHz).

For some low-redshift TDFs with powerful jets we can resolve the new radio source. With ngVLA  observations in Band 4 ($\approx 30$~GHz, yielding a maximum resolutoin of a few mas) we can likely resolve jets similar to Swift~J1644+57 at $z<0.1$ within one year after launch. The number of detections at this redshift is low ($<1$ per year); including VLBI to the ngVLA would signficantly increase the detection rate. Resolving a newly created jet presents a unique oppurtunity, allowing a measurement of the magnetic field of the radio-emitting region without using the equipartition assumption \citep[e.g.,][]{Falcke99}. Furthermore, by extrapolating the source size measurements backwards in time, we obtain the launch date of the jet. This launch date can be compared to the light curve of the TDF disk emission, providing a direct test of the prediction that jet launching requires super-Eddington accretion rates \citep[e.g.,][]{Tchekhovskoy13}. 

Comparing the rate of tidal disruption jets detected in ngVLA surveys to the rate of TDFs detected by their thermal disk emission (in X-ray or optical surveys) we obtain the jet efficiency (i.e., the fraction of disruptions that lead to jets). Using multi-frequency radio light curves we can measure jet efficiency as a function of the total jet energy. Within the accretion paradigm we would expect that the efficiency is constant because the jet energy reflects the amount of stellar debris that was accreted. If, on the other hand, jets are powered by spin, we would expect a few TDFs with a total jet energy that exceeds the rest-mass energy of the accreted debris. 


In this chapter we focused on using radio observations of TDFs to constrain models of jet formation. Yet the radio light curves of these sources can also be used as tools to measure the gas density in nuclei of their host galaxies \citep{Zauderer11,Berger12,Generozov17,Eftekhari18}. Since we can detect sources similar to Swift~J1644+57 to very high redshifts (Figure~\ref{fig:radiolc}), we can use ngVLA observations to probe the environments of black holes in the earliest stages of their growth.


\bibliography{$HOME/Documents/articles/general_desk.bib}  

\begin{thebibliography}{}
\expandafter\ifx\csname natexlab\endcsname\relax\def\natexlab#1{#1}\fi

\bibitem[{{Alexander} {et~al.}(2016){Alexander}, {Berger}, {Guillochon},
  {Zauderer}, \& {Williams}}]{Alexander16}
{Alexander}, K.~D., {Berger}, E., {Guillochon}, J., {Zauderer}, B.~A., \&
  {Williams}, P.~K.~G. 2016, \apjl, 819, L25

\bibitem[{{Alexander} {et~al.}(2017){Alexander}, {Wieringa}, {Berger},
  {Saxton}, \& {Komossa}}]{Alexander17}
{Alexander}, K.~D., {Wieringa}, M.~H., {Berger}, E., {Saxton}, R.~D., \&
  {Komossa}, S. 2017, \apj, 837, 153

\bibitem[{{Arcavi} {et~al.}(2014){Arcavi}, {Gal-Yam}, {Sullivan}, {Pan},
  {Cenko}, {Horesh}, {Ofek}, {De Cia}, {Yan}, {Yang}, {Howell}, {Tal},
  {Kulkarni}, {Tendulkar}, {Tang}, {Xu}, {Sternberg}, {Cohen}, {Bloom},
  {Nugent}, {Kasliwal}, {Perley}, {Quimby}, {Miller}, {Theissen}, \&
  {Laher}}]{Arcavi14}
{Arcavi}, I., {Gal-Yam}, A., {Sullivan}, M., {et~al.} 2014, \apj, 793, 38

\bibitem[{{Berger} {et~al.}(2012){Berger}, {Zauderer}, {Pooley}, {Soderberg},
  {Sari}, {Brunthaler}, \& {Bietenholz}}]{Berger12}
{Berger}, E., {Zauderer}, A., {Pooley}, G.~G., {et~al.} 2012, \apj, 748, 36

\bibitem[{{Blandford} \& {Znajek}(1977)}]{BlandfordZnajek77}
{Blandford}, R.~D., \& {Znajek}, R.~L. 1977, \mnras, 179, 433

\bibitem[{{Bloom} {et~al.}(2011){Bloom}, {Giannios}, {Metzger}, {Cenko},
  {Perley}, {Butler}, {Tanvir}, {Levan}, {O'Brien}, {Strubbe}, {De Colle},
  {Ramirez-Ruiz}, {Lee}, {Nayakshin}, {Quataert}, {King}, {Cucchiara},
  {Guillochon}, {Bower}, {Fruchter}, {Morgan}, \& {van der Horst}}]{Bloom11}
{Bloom}, J.~S., {Giannios}, D., {Metzger}, B.~D., {et~al.} 2011, Science, 333,
  203

\bibitem[{{Bower} {et~al.}(2013){Bower}, {Metzger}, {Cenko}, {Silverman}, \&
  {Bloom}}]{Bower13}
{Bower}, G.~C., {Metzger}, B.~D., {Cenko}, S.~B., {Silverman}, J.~M., \&
  {Bloom}, J.~S. 2013, \apj, 763, 84

\bibitem[{{Brown} {et~al.}(2015){Brown}, {Levan}, {Stanway}, {Tanvir}, {Cenko},
  {Berger}, {Chornock}, \& {Cucchiaria}}]{Brown15}
{Brown}, G.~C., {Levan}, A.~J., {Stanway}, E.~R., {et~al.} 2015, \mnras, 452,
  4297

\bibitem[{{Burrows} {et~al.}(2011){Burrows}, {Kennea}, {Ghisellini}, {Mangano},
  {Zhang}, {Page}, {Eracleous}, {Romano}, {Sakamoto}, {Falcone}, {Osborne},
  {Campana}, {Beardmore}, {Breeveld}, {Chester}, {Corbet}, {Covino},
  {Cummings}, {D'Avanzo}, {D'Elia}, {Esposito}, {Evans}, {Fugazza}, {Gelbord},
  {Hiroi}, {Holland}, {Huang}, {Im}, {Israel}, {Jeon}, {Jeon}, {Jun}, {Kawai},
  {Kim}, {Krimm}, {Marshall}, {P.~M{\'e}sz{\'a}ros}, {Negoro}, {Omodei},
  {Park}, {Perkins}, {Sugizaki}, {Sung}, {Tagliaferri}, {Troja}, {Ueda},
  {Urata}, {Usui}, {Antonelli}, {Barthelmy}, {Cusumano}, {Giommi}, {Melandri},
  {Perri}, {Racusin}, {Sbarufatti}, {Siegel}, \& {Gehrels}}]{Burrows11}
{Burrows}, D.~N., {Kennea}, J.~A., {Ghisellini}, G., {et~al.} 2011, \nat, 476,
  421

\bibitem[{{Cenko} {et~al.}(2012){Cenko}, {Bloom}, {Kulkarni}, {Strubbe},
  {Miller}, {Butler}, {Quimby}, {Gal-Yam}, {Ofek}, {Quataert}, {Bildsten},
  {Poznanski}, {Perley}, {Morgan}, {Filippenko}, {Frail}, {Arcavi}, {Ben-Ami},
  {Cucchiara}, {Fassnacht}, {Green}, {Hook}, {Howell}, {Lagattuta}, {Law},
  {Kasliwal}, {Nugent}, {Silverman}, {Sullivan}, {Tendulkar}, \&
  {Yaron}}]{Cenko12}
{Cenko}, S.~B., {Bloom}, J.~S., {Kulkarni}, S.~R., {et~al.} 2012, \mnras, 420,
  2684

\bibitem[{{Dong} {et~al.}(2016){Dong}, {Shappee}, {Prieto}, {Jha}, {Stanek},
  {Holoien}, {Kochanek}, {Thompson}, {Morrell}, {Thompson}, {Basu}, {Beacom},
  {Bersier}, {Brimacombe}, {Brown}, {Bufano}, {Chen}, {Conseil}, {Danilet},
  {Falco}, {Grupe}, {Kiyota}, {Masi}, {Nicholls}, {Olivares E.}, {Pignata},
  {Pojmanski}, {Simonian}, {Szczygiel}, \& {Wo{\'z}niak}}]{Dong16}
{Dong}, S., {Shappee}, B.~J., {Prieto}, J.~L., {et~al.} 2016, Science, 351, 257

\bibitem[{{Donnarumma} \& {Rossi}(2015)}]{Donnarumma15}
{Donnarumma}, I., \& {Rossi}, E.~M. 2015, \apj, 803, 36

\bibitem[{{Donnarumma} {et~al.}(2015){Donnarumma}, {Rossi}, {Fender},
  {Komossa}, {Paragi}, {Van Velzen}, \& {Prandoni}}]{Donnarumma15a}
{Donnarumma}, I., {Rossi}, E.~M., {Fender}, R., {et~al.} 2015, Advancing
  Astrophysics with the Square Kilometre Array (AASKA14), 54

\bibitem[{{Eftekhari} {et~al.}(2018){Eftekhari}, {Berger}, {Zauderer},
  {Margutti}, \& {Alexander}}]{Eftekhari18}
{Eftekhari}, T., {Berger}, E., {Zauderer}, B.~A., {Margutti}, R., \&
  {Alexander}, K.~D. 2018, \apj, 854, 86

\bibitem[{{Evans} \& {Kochanek}(1989)}]{EvansKochanek89}
{Evans}, C.~R., \& {Kochanek}, C.~S. 1989, \apjl, 346, L13

\bibitem[{{Falcke} {et~al.}(2004){Falcke}, {K{\"o}rding}, \&
  {Markoff}}]{Falcke04}
{Falcke}, H., {K{\"o}rding}, E., \& {Markoff}, S. 2004, \aap, 414, 895

\bibitem[{{Falcke} {et~al.}(1999){Falcke}, {Bower}, {Lobanov}, {Krichbaum},
  {Patnaik}, {Aller}, {Aller}, {Ter{\"a}sranta}, {Wright}, \&
  {Sandell}}]{Falcke99}
{Falcke}, H., {Bower}, G.~C., {Lobanov}, A.~P., {et~al.} 1999, \apjl, 514, L17

\bibitem[{Fender {et~al.}(2017)Fender, Woudt, Armstrong, Groot, McBride,
  Miller-Jones, Mooley, Stappers, Wijers, Bietenholz, Blyth, Bottcher, Buckley,
  Charles, Chomiuk, Coppejans, Corbel, Coriat, Daigne, de~Blok, Falcke, Girard,
  Heywood, Horesh, Horrell, Jonker, Joseph, Kamble, Knigge, Koerding, Kotze,
  Kouveliotou, Lynch, Maccarone, Meintjes, Migliari, Murphy, Nagayama,
  Nelemans, Nicholson, O'Brien, Oodendaal, Oozeer, Osborne, Perez-Torres,
  Ratcliffe, Ribeiro, Rol, Rushton, Scaife, Schurch, Sivakoff, Staley, Steeghs,
  Stewart, Swinbank, van~der Heyden, van~der Horst, van Soelen, Vergani,
  Warner, \& Wiersema}]{Fender17}
Fender, R., Woudt, P., Armstrong, R., {et~al.} 2017, arXiv:1711.04132

\bibitem[{{Fender} {et~al.}(2004){Fender}, {Belloni}, \& {Gallo}}]{Fender04}
{Fender}, R.~P., {Belloni}, T.~M., \& {Gallo}, E. 2004, \mnras, 355, 1105

\bibitem[{{Generozov} {et~al.}(2017){Generozov}, {Mimica}, {Metzger}, {Stone},
  {Giannios}, \& {Aloy}}]{Generozov17}
{Generozov}, A., {Mimica}, P., {Metzger}, B.~D., {et~al.} 2017, \mnras, 464,
  2481

\bibitem[{{Gezari} {et~al.}(2009){Gezari}, {Heckman}, {Cenko}, {Eracleous},
  {Forster}, {Gon{\c c}alves}, {Martin}, {Morrissey}, {Neff}, {Seibert},
  {Schiminovich}, \& {Wyder}}]{Gezari09}
{Gezari}, S., {Heckman}, T., {Cenko}, S.~B., {et~al.} 2009, \apj, 698, 1367

\bibitem[{{Giannios} \& {Metzger}(2011)}]{Giannios11}
{Giannios}, D., \& {Metzger}, B.~D. 2011, \mnras, 416, 2102

\bibitem[{{Graham} {et~al.}(2017){Graham}, {Djorgovski}, {Drake}, {Stern},
  {Mahabal}, {Glikman}, {Larson}, \& {Christensen}}]{Graham17}
{Graham}, M.~J., {Djorgovski}, S.~G., {Drake}, A.~J., {et~al.} 2017, \mnras,
  470, 4112

\bibitem[{{Hills}(1975)}]{Hills75}
{Hills}, J.~G. 1975, \nat, 254, 295

\bibitem[{{Holoien} {et~al.}(2016){Holoien}, {Kochanek}, {Prieto}, {Stanek},
  {Dong}, {Shappee}, {Grupe}, {Brown}, {Basu}, {Beacom}, {Bersier},
  {Brimacombe}, {Danilet}, {Falco}, {Guo}, {Jose}, {Herczeg}, {Long},
  {Pojmanski}, {Simonian}, {Szczygie{\l}}, {Thompson}, {Thorstensen}, {Wagner},
  \& {Wo{\'z}niak}}]{Holoien16}
{Holoien}, T.~W.-S., {Kochanek}, C.~S., {Prieto}, J.~L., {et~al.} 2016, \mnras,
  455, 2918

\bibitem[{{Kara} {et~al.}(2016){Kara}, {Miller}, {Reynolds}, \& {Dai}}]{Kara16}
{Kara}, E., {Miller}, J.~M., {Reynolds}, C., \& {Dai}, L. 2016, \nat, 535, 388

\bibitem[{{Kellermann} {et~al.}(1989){Kellermann}, {Sramek}, {Schmidt},
  {Shaffer}, \& {Green}}]{Kellermann89}
{Kellermann}, K.~I., {Sramek}, R., {Schmidt}, M., {Shaffer}, D.~B., \& {Green},
  R. 1989, \aj, 98, 1195

\bibitem[{{Koay} {et~al.}(2016){Koay}, {Vestergaard}, {Bignall}, {Reynolds}, \&
  {Peterson}}]{Koay16}
{Koay}, J.~Y., {Vestergaard}, M., {Bignall}, H.~E., {Reynolds}, C., \&
  {Peterson}, B.~M. 2016, \mnras, 460, 304

\bibitem[{{Komossa}(2015)}]{Komossa15}
{Komossa}, S. 2015, Journal of High Energy Astrophysics, 7, 148

\bibitem[{{K{\"o}rding} {et~al.}(2006){K{\"o}rding}, {Jester}, \&
  {Fender}}]{Koerding06}
{K{\"o}rding}, E.~G., {Jester}, S., \& {Fender}, R. 2006, \mnras, 372, 1366

\bibitem[{{Krolik} {et~al.}(2016){Krolik}, {Piran}, {Svirski}, \&
  {Cheng}}]{Krolik16}
{Krolik}, J., {Piran}, T., {Svirski}, G., \& {Cheng}, R.~M. 2016, \apj, 827,
  127

\bibitem[{{Kr{\"u}hler} {et~al.}(2018){Kr{\"u}hler}, {Fraser}, {Leloudas},
  {Schulze}, {Stone}, {van Velzen}, {Amorin}, {Hjorth}, {Jonker}, {Kann},
  {Kim}, {Kuncarayakti}, {Mehner}, \& {Nicuesa Guelbenzu}}]{Kruhler18}
{Kr{\"u}hler}, T., {Fraser}, M., {Leloudas}, G., {et~al.} 2018, \aap, 610, A14

\bibitem[{{LaMassa} {et~al.}(2015){LaMassa}, {Cales}, {Moran}, {Myers},
  {Richards}, {Eracleous}, {Heckman}, {Gallo}, \& {Urry}}]{LaMassa15}
{LaMassa}, S.~M., {Cales}, S., {Moran}, E.~C., {et~al.} 2015, \apj, 800, 144

\bibitem[{{Leloudas} {et~al.}(2016){Leloudas}, {Fraser}, {Stone}, {van Velzen},
  {Jonker}, {Arcavi}, {Fremling}, {Maund}, {Smartt}, {Kr{\`\i}hler},
  {Miller-Jones}, {Vreeswijk}, {Gal-Yam}, {Mazzali}, {De Cia}, {Howell},
  {Inserra}, {Patat}, {de Ugarte Postigo}, {Yaron}, {Ashall}, {Bar},
  {Campbell}, {Chen}, {Childress}, {Elias-Rosa}, {Harmanen}, {Hosseinzadeh},
  {Johansson}, {Kangas}, {Kankare}, {Kim}, {Kuncarayakti}, {Lyman}, {Magee},
  {Maguire}, {Malesani}, {Mattila}, {McCully}, {Nicholl}, {Prentice},
  {Romero-Ca{\~n}izales}, {Schulze}, {Smith}, {Sollerman}, {Sullivan},
  {Tucker}, {Valenti}, {Wheeler}, \& {Young}}]{Leloudas16}
{Leloudas}, G., {Fraser}, M., {Stone}, N.~C., {et~al.} 2016, Nature Astronomy,
  1, 0002

\bibitem[{{Levan} {et~al.}(2011){Levan}, {Tanvir}, {Cenko}, {Perley},
  {Wiersema}, {Bloom}, {Fruchter}, {Postigo}, {O'Brien}, {Butler}, {van der
  Horst}, {Leloudas}, {Morgan}, {Misra}, {Bower}, {Farihi}, {Tunnicliffe},
  {Modjaz}, {Silverman}, {Hjorth}, {Th{\"o}ne}, {Cucchiara}, {Cer{\'o}n},
  {Castro-Tirado}, {Arnold}, {Bremer}, {Brodie}, {Carroll}, {Cooper}, {Curran},
  {Cutri}, {Ehle}, {Forbes}, {Fynbo}, {Gorosabel}, {Graham}, {Hoffman},
  {Guziy}, {Jakobsson}, {Kamble}, {Kerr}, {Kasliwal}, {Kouveliotou},
  {Kocevski}, {Law}, {Nugent}, {Ofek}, {Poznanski}, {Quimby}, {Rol},
  {Romanowsky}, {S{\'a}nchez-Ram{\'{\i}}rez}, {Schulze}, {Singh}, {van
  Spaandonk}, {Starling}, {Strom}, {Tello}, {Vaduvescu}, {Wheatley}, {Wijers},
  {Winters}, \& {Xu}}]{Levan11}
{Levan}, A.~J., {Tanvir}, N.~R., {Cenko}, S.~B., {et~al.} 2011, Science, 333,
  199

\bibitem[{{Levan} {et~al.}(2016){Levan}, {Tanvir}, {Brown}, {Metzger}, {Page},
  {Cenko}, {O'Brien}, {Lyman}, {Wiersema}, {Stanway}, {Fruchter}, {Perley}, \&
  {Bloom}}]{Levan16}
{Levan}, A.~J., {Tanvir}, N.~R., {Brown}, G.~C., {et~al.} 2016, \apj, 819, 51

\bibitem[{{Margutti} {et~al.}(2017){Margutti}, {Metzger}, {Chornock},
  {Milisavljevic}, {Berger}, {Blanchard}, {Guidorzi}, {Migliori}, {Kamble},
  {Lunnan}, {Nicholl}, {Coppejans}, {Dall'Osso}, {Drout}, {Perna}, \&
  {Sbarufatti}}]{Margutti17}
{Margutti}, R., {Metzger}, B.~D., {Chornock}, R., {et~al.} 2017, \apj, 836, 25

\bibitem[{{Metzger} {et~al.}(2015){Metzger}, {Williams}, \&
  {Berger}}]{Metzger15}
{Metzger}, B.~D., {Williams}, P.~K.~G., \& {Berger}, E. 2015, \apj, 806, 224

\bibitem[{{Murphy} {et~al.}(2013){Murphy}, {Chatterjee}, {Kaplan}, {Banyer},
  {Bell}, {Bignall}, {Bower}, {Cameron}, {Coward}, {Cordes}, {Croft}, {Curran},
  {Djorgovski}, {Farrell}, {Frail}, {Gaensler}, {Galloway}, {Gendre}, {Green},
  {Hancock}, {Johnston}, {Kamble}, {Law}, {Lazio}, {Lo}, {Macquart}, {Rea},
  {Rebbapragada}, {Reynolds}, {Ryder}, {Schmidt}, {Soria}, {Stairs}, {Tingay},
  {Torkelsson}, {Wagstaff}, {Walker}, {Wayth}, \& {Williams}}]{Murphy13}
{Murphy}, T., {Chatterjee}, S., {Kaplan}, D.~L., {et~al.} 2013, \pasa, 30, e006

\bibitem[{{Pasham} \& {van Velzen}(2018)}]{PashamvanVelzen17}
{Pasham}, D.~R., \& {van Velzen}, S. 2018, ApJ, 856, 14

\bibitem[{{Pasham} {et~al.}(2015){Pasham}, {Cenko}, {Levan}, {Bower}, {Horesh},
  {Brown}, {Dolan}, {Wiersema}, {Filippenko}, {Fruchter}, {Greiner}, {O'Brien},
  {Page}, {Rau}, \& {Tanvir}}]{Pasham15}
{Pasham}, D.~R., {Cenko}, S.~B., {Levan}, A.~J., {et~al.} 2015, \apj, 805, 68

\bibitem[{{Rees}(1988)}]{Rees88}
{Rees}, M.~J. 1988, \nat, 333, 523

\bibitem[{{Stone} {et~al.}(2018){Stone}, {Kesden}, {Cheng}, \&
  {Velzen}}]{Stone18}
{Stone}, N.~C., {Kesden}, M., {Cheng}, R.~M., \& {Velzen}, S.~v. 2018,
  arXiv:1801.10180v1

\bibitem[{{Tchekhovskoy} {et~al.}(2014){Tchekhovskoy}, {Metzger}, {Giannios},
  \& {Kelley}}]{Tchekhovskoy13}
{Tchekhovskoy}, A., {Metzger}, B.~D., {Giannios}, D., \& {Kelley}, L.~Z. 2014,
  \mnras, 437, 2744

\bibitem[{{van Velzen}(2018)}]{vanVelzen18}
{van Velzen}, S. 2018, \apj, 852, 72

\bibitem[{{van Velzen} {et~al.}(2013){van Velzen}, {Frail}, {K{\"o}rding}, \&
  {Falcke}}]{vanVelzen12b}
{van Velzen}, S., {Frail}, D.~A., {K{\"o}rding}, E., \& {Falcke}, H. 2013,
  \aap, 552, A5

\bibitem[{van Velzen {et~al.}(2011)van Velzen, K\"ording, \&
  Falcke}]{vanVelzen11}
van Velzen, S., K\"ording, E., \& Falcke, H. 2011, \mnras, 417, L51

\bibitem[{{van Velzen} {et~al.}(2011){van Velzen}, {Farrar}, {Gezari},
  {Morrell}, {Zaritsky}, {{\"O}stman}, {Smith}, {Gelfand}, \&
  {Drake}}]{vanVelzen10}
{van Velzen}, S., {Farrar}, G.~R., {Gezari}, S., {et~al.} 2011, \apj, 741, 73

\bibitem[{{van Velzen} {et~al.}(2016){van Velzen}, {Anderson}, {Stone},
  {Fraser}, {Wevers}, {Metzger}, {Jonker}, {van der Horst}, {Staley}, {Mendez},
  {Miller-Jones}, {Hodgkin}, {Campbell}, \& {Fender}}]{vanVelzen16}
{van Velzen}, S., {Anderson}, G.~E., {Stone}, N.~C., {et~al.} 2016, Science,
  351, 62

\bibitem[{{Vogelsberger} {et~al.}(2014){Vogelsberger}, {Genel}, {Springel},
  {Torrey}, {Sijacki}, {Xu}, {Snyder}, {Bird}, {Nelson}, \&
  {Hernquist}}]{Vogelsberger14}
{Vogelsberger}, M., {Genel}, S., {Springel}, V., {et~al.} 2014, \nat, 509, 177

\bibitem[{{Zauderer} {et~al.}(2013){Zauderer}, {Berger}, {Margutti}, {Pooley},
  {Sari}, {Soderberg}, {Brunthaler}, \& {Bietenholz}}]{Zauderer13}
{Zauderer}, B.~A., {Berger}, E., {Margutti}, R., {et~al.} 2013, \apj, 767, 152

\bibitem[{{Zauderer} {et~al.}(2011){Zauderer}, {Berger}, {Soderberg}, {Loeb},
  {Narayan}, {Frail}, {Petitpas}, {Brunthaler}, {Chornock}, {Carpenter},
  {Pooley}, {Mooley}, {Kulkarni}, {Margutti}, {Fox}, {Nakar}, {Patel},
  {Volgenau}, {Culverhouse}, {Bietenholz}, {Rupen}, {Max-Moerbeck}, {Readhead},
  {Richards}, {Shepherd}, {Storm}, \& {Hull}}]{Zauderer11}
{Zauderer}, B.~A., {Berger}, E., {Soderberg}, A.~M., {et~al.} 2011, \nat, 476,
  425

\end{thebibliography}

\end{document}